\documentclass{PoS}
\usepackage{dsfont}

\title{Density of states approach for lattice field theory with topological terms\thanks{Based on a parallel talk by 
C.~Gattringer.}}

\ShortTitle{Density of states approach for lattice field theory with topological terms}

\author{Christof Gattringer$^1$\footnote{Currently on leave of absence from University of Graz, 8010 Graz, Austria.} , 
Oliver Orasch$^2$\\

\vskip2mm

$^1\;$FWF - Austrian Science Fund, 1090 Vienna, Austria 

$^2\;$Institute of Physics\footnote{Member of NAWI Graz.} , University of Graz, 8010 Graz, Austria

\vskip2mm
        
E-mail: \email{christof.gattringer@fwf.ac.at} \\
E-mail: \email{oliver.orasch@uni-graz.at}  }

\abstract{We discuss a new density of states (DoS) approach to solve the complex action problem that is caused by 
topological terms. The key ingredient is to use open boundary conditions for (at least) one of the directions, such that the
quantization of the topological charge is lifted and the density becomes a regular function. We employ the DoS FFA method 
and compute the density of states as a function of the topological charge. Subsequent integration with 
suitable factors gives rise to the observables we are interested in. We here explore two test cases: U(1) lattice gauge 
theory in two dimensions, and SU(2) lattice gauge theory in four dimensions. Since the 2-d case has an exact solution we
may use it to assess the method, in particular to establish the equivalence of the open boundary results with the usual choice of 
periodic boundary conditions. The SU(2) case is a first step of developing the techniques towards their eventual 
application in full QCD.}

\FullConference{38th International Symposium on Lattice Field Theory - Lattice2021\\
		July 26 -  30, 2021\\
		Zoom/Gather@MIT}

\begin{document}

\section{Introductory comments}

\noindent
When adding a topological term the gauge field action acquires an imaginary part such 
that the Boltzmann factor of the corresponding Euclidean 
path integral becomes complex and thus cannot be used as a weight factor for a Monte Carlo calculation in the lattice formulation. 
In principle, density of states techniques -- initially introduced in lattice field theory in \cite{Gocksch1,Gocksch2} -- 
are a possible approach 
and, based on various recent technological developments, were applied in a wide range of applications with complex action problems
\cite{Schmidt:2005ap}  -- \cite{Gattringer:2020mbf}. However,
the case of treating the complex action problem emerging from a topological term is somewhat subtle, because in the usual 
formulation with periodic boundary conditions the topological charge becomes quantized to integers in the continuum 
limit such that the density will approach a superposition of Dirac deltas (see, e.g., \cite{dos_theta}).
If a geometrical or a fermionic definition is used the topological charge is quantized also at finite lattice constant, resulting 
in a superposition of Dirac deltas also at finite lattice spacing. As a consequence the density is hard to access with the new 
DoS methods.

It is well known that using open boundary conditions lifts the quantization of the topological charge and in the thermodynamic 
and continuum limits describes the correct physics. Using open boundary conditions in lattice simulations was proposed in 
\cite{openbc1,openbc2} as a tool for reducing topological autocorrelation, which, due to the index theorem, 
is a problem that is related to our
problem of singular densities of states when a topological term is added. In \cite{Gattringer:2020mbf} it was demonstrated, that 
also the 
latter problem can be solved with open boundary conditions and that for topological quantities the correct continuum limit 
is found (see also \cite{pushan}). In this contribution we review the new techniques, discuss the case of 
2-d U(1) lattice gauge theory with a $\theta$-term as
an example where analytic reference results can be used to assess the approach, and finally present first tests for 4-d
SU(2) lattice gauge theory with a $\theta$-term.

\section{Density of states formalism for lattice gauge theories with a $\theta$-term}

\noindent
The generic form for vacuum expectation values in pure gauge theory with a $\theta$-term is given by
\begin{equation}
\langle O \rangle_\theta \, = \, \frac{1}{Z_\theta} \! 
\int \!\! D[A] \, e^{\, - \, S[A] \;  -  \; i \theta \, Q[A]} \, O[A] 
\qquad  \mbox{with} \qquad 
Z_\theta \, = \int \!\! D[A] \, e^{\, - \, S[A] \; - \; i \theta \, Q[A]} \; ,
\end{equation}
where $S[A]$ denotes the gauge field action and $Q[A]$ the topological charge. We assume that the theory is already regularized on a lattice such that $S[A]$ and $Q[A]$ are suitable discretizations of their continuum counterparts and $D[A]$ denotes the usual 
product of Haar measures for the link variables. 

For non-vanishing topological angle $\theta$
the Boltzmann factor $e^{\, - \, S[A] \;  -  \; i \theta \, Q[A]}$ is complex and thus cannot be used as a probability in a Monte Carlo 
simulation. For setting up the density of states approach we introduce (generalized) densities
\begin{equation}
\rho^{(J)}(x) \; = \, \int \!\! D[A] \; 
e^{\, - \, S[A]} \; J[A] \; 
\delta\Big(x - Q[A]\Big) \; ,
\label{densities}
\end{equation}
where we allow for the insertion of some functional $J[A]$ of the gauge fields. 

Obviously, when $Q[A]$ becomes integer in the continuum limit (or is integer also at finite lattice spacing if a geometrical 
or a fermionic definition of $Q[A]$ is used) the density will be the aforementioned superposition of Dirac deltas.  
However, we may use open boundary conditions to lift 
the quantization of the topological charge. Thus for open boundary conditions 
in at least one of the lattice directions the density remains smooth also in the continuum limit. This is a key step in the 
approach presented here. 

Using the densities (\ref{densities}) we may write the vacuum expectation values as 
\begin{equation}
\langle O \rangle_\theta \, = \, \frac{1}{Z_\theta} \! 
\int \! dx \; \rho^{(O)}(x) \; 
e^{ \, - \, i  \, \theta \, x} 
\qquad  \mbox{with} \qquad 
Z_\theta \, = \int \! dx \; \rho^{(\mathds{1})}(x) \; 
e^{ \, - \, i \, \theta \, x} \; .
\label{integrals}
\end{equation} 
\begin{figure}[t] 
   \includegraphics[width=65mm,clip]{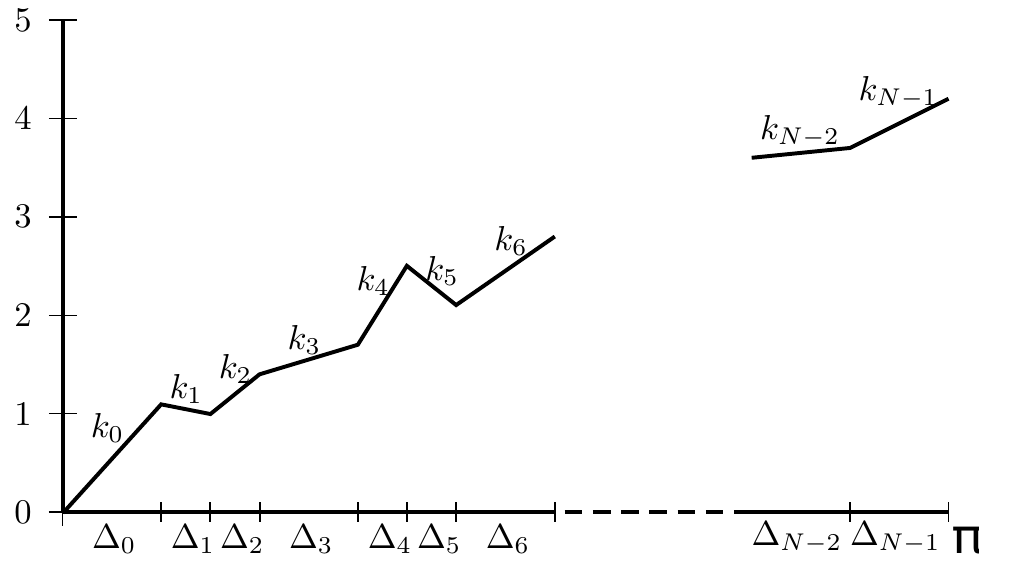} 
   \hspace{4mm}
   \includegraphics[width=80mm,clip]{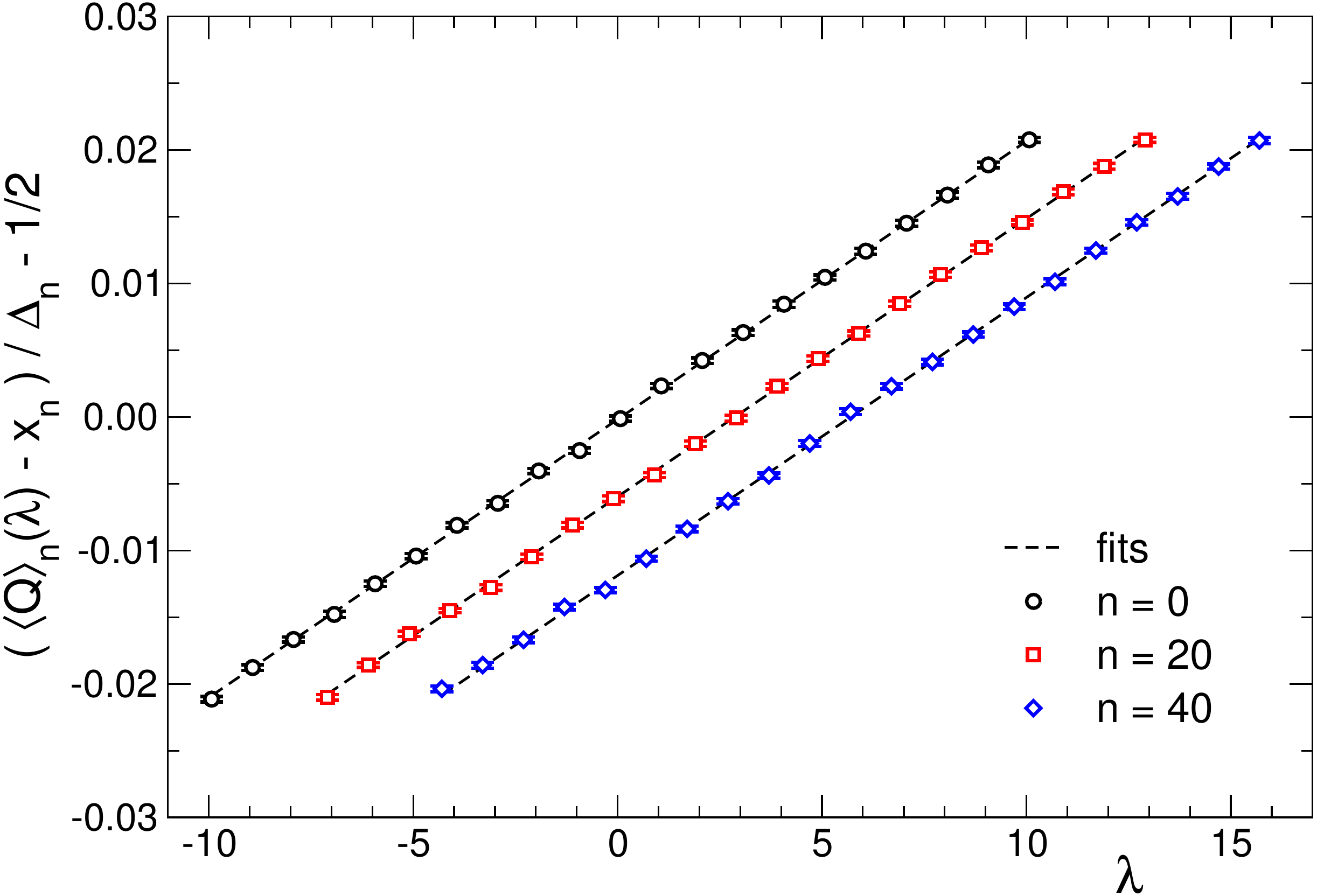}
   \caption{Lhs.: Parameterization of the exponent of the density as a continuous and piecewise linear function. Rhs.: Fit of the 
   rescaled restricted vacuum expectation values (figure reproduced from \cite{Gattringer:2020mbf}).}
   \label{figure_1}
\end{figure}
For an actual determination of the densities we need to parameterize them in a suitable way. As a first step towards such a 
parameterization we truncate the range of the argument $x$ to the interval $[0,l_{max}]$. Such a truncation is justified since the 
densities are either even or odd and decay fast for large arguments $x$, a property that of course needs to be checked in the end. 
Next we divide the (truncated) $x$-range $[0,l_{max}]$ into $N$ intervals $I_n, \; n = 0, 1 \, ... \, N-1$ of sizes $\Delta_n$.
For the densities we now make the ansatz:
\begin{equation}
 \rho(x) \; = \; e^{ \, - \, L(x) } \quad \mbox{with} \; 
 L(x) \; \mbox{continuous and piecewise linear on the intervals} \; \; I_n \; .
\end{equation} 
The parameterization of the exponent of the density with the piecewise linear and continuous function is illustrated in 
the lhs.\ plot of Fig.~\ref{figure_1}.
Imposing the normalization $\rho(0) = 1$ completely determines the densities
$\rho(x)$ in terms of the slopes $k_n$ defined for each of the intervals $I_n$,
\begin{equation}
\rho(x) \; = \;  A_n \; e^{ \, -  \, x \, k_n} \qquad \mbox{for}  \quad 
x \in I_n 
\qquad \mbox{with}  \quad 
A_n \; = \; e^{\, -  \sum_{j=0}^{n-1} \big[ k_j - k_n \big] \Delta_j } \; .
\label{paramdens}
\end{equation} 
To determine the slopes we use so-called restricted vacuum expectation values \cite{Langfeld:2012ah}--\cite{Langfeld:2015fua}
which are defined as
\begin{equation}
\langle Q \rangle_n(\lambda) \; = \; 
\frac{1}{Z_n(\lambda)}  
\int \!\! D[A] \, e^{\, - \, S[A]} \,
J[A] \;
e^{ \, \lambda \, Q[A]} \; Q[A]  \; \Theta_n\Big( Q[A] \Big) 
\quad \mbox{with} \quad 
\Theta_n(x) \; = \; \left\{ 
\begin{array}{cc}
1 & \mbox{for} \;\;\; x \in I_n \\
0 & \mbox{for} \;\;\; x \notin I_n 
\end{array}
\right. .
\end{equation}
Obviously the function $\Theta_n\Big( Q[A] \Big)$ restricts the path integral such that the values of 
the topological charge are required to be in the interval $I_n$. The restricted vacuum expectation values 
$\langle Q \rangle_n(\lambda)$ are free of the complex action problem and thus can be evaluated with 
standard Monte Carlo simulations. We will compute them as a function of the parameter $\lambda \in \mathds{R}\,$. 

However, the $\langle Q \rangle_n(\lambda)$ may also be computed from the parameterized density 
$\rho(x) \; = \;  A_n \; e^{ \, -  \, x \, k_n}$ of Eq.~(\ref{paramdens}). A trivial calculation gives
\begin{equation}
\langle Q \rangle_n(\lambda) \; = \; \frac{d}{d\, \lambda}  \ln \int_{x_n}^{x_{n+1}} \!\! dx \; \rho(x) \; e^{\, \lambda \, x}
\; = \; g_n (\lambda- k_n) \; ,
\end{equation}
with some function $g_n$ (see, e.g., \cite{FFA_2} for details). After an additive and a multiplicative 
normalization of the restricted vacuum expectation values
one finds the more convenient form
\begin{equation}
\frac{ \langle \, Q \, \rangle_n(\lambda) - x_n}{\Delta_n}
- \frac{1}{2} \, = \, h\Big(\Delta_n\big[\lambda-k_n\big]\Big) \; ,
\label{Vdef}
\end{equation}
where the function $h(s)$ is given by
\begin{equation} 
h(s) \, \equiv \,  \frac{1}{1-e^{-s}}-\frac{1}{s}-\frac{1}{2}  \; .
\label{hdef}
\end{equation}
We thus may fit the Monte Carlo data for 
$(\langle Q  \rangle_n(\lambda) - x_n) / \Delta_n - 1/2$ with the function $h(\Delta_n[\lambda-k_n])$ and in this 
way determine the slopes $k_n$ from simple stable 1-parameter fits (DoS FFA approach \cite{Z3_FFA_1} -- \cite{FFA_3}). 
The rhs.\ plot of Fig.~\ref{figure_1} provides 
an example of such fits, where we show the results for different intervals $I_n$ for $n = 0, 20$ and 40. The data are for the simulation 
of 2-d U(1) lattice gauge theory presented in \cite{Gattringer:2020mbf}. 

\section{Tests in U(1) lattice gauge theory}

\noindent
The idea of applying DoS methods with open boundary conditions for treating the complex action problem from a topological term 
was initially tested for U(1) lattice gauge theory in 2-d where exact results can be used to evaluate the approach  
\cite{Gattringer:2020mbf}. Using dual variables 
these exact results may be obtained for both, open and periodic boundary conditions, such that 
one can compare the correct continuum limit for both types of boundary conditions and in \cite{Gattringer:2020mbf} it was shown that
indeed both choices give rise to the correct continuum limit. 

\begin{figure}[t] 
\includegraphics[scale=0.302,clip]{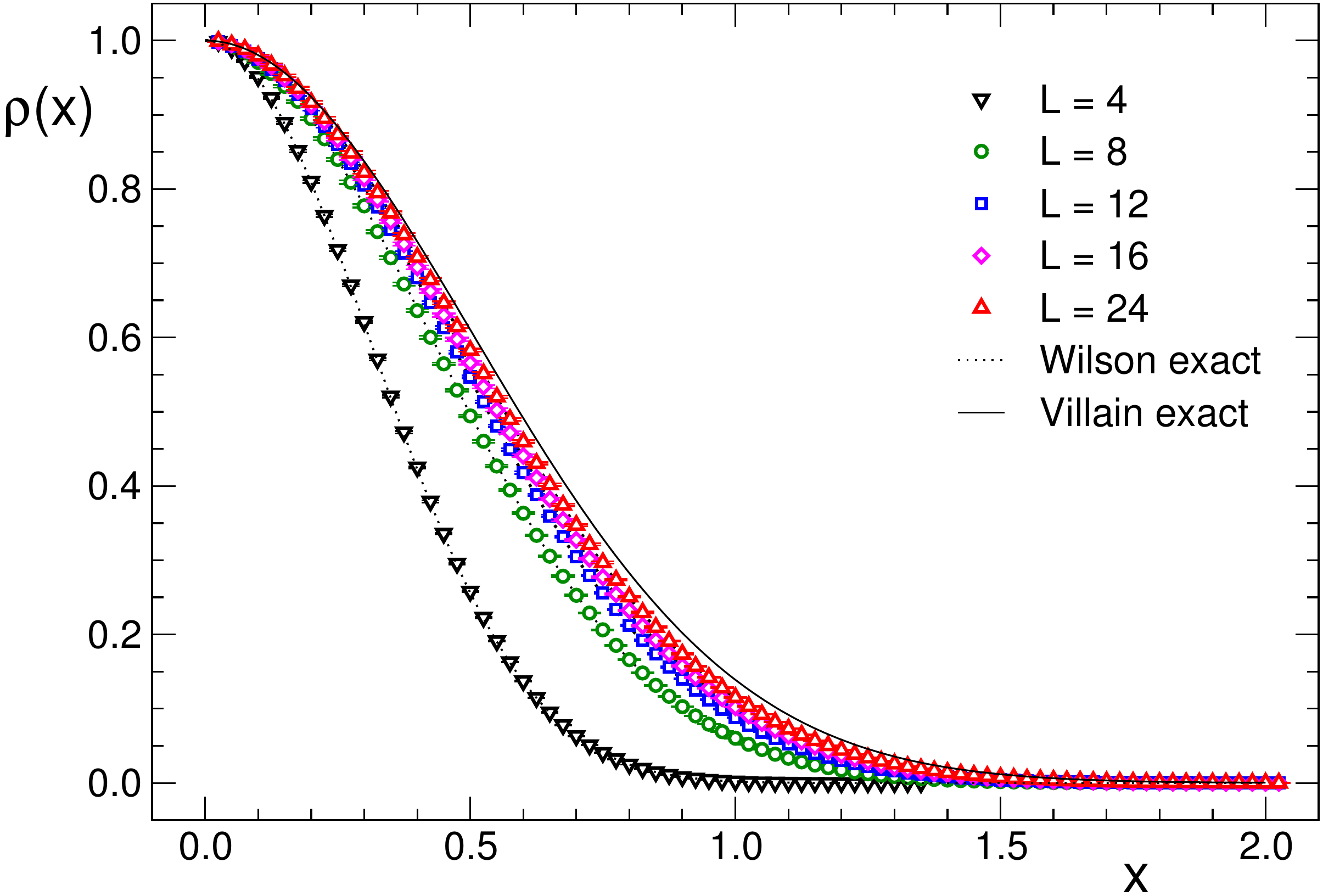} 
\hspace{4mm}
\includegraphics[scale=0.30,clip]{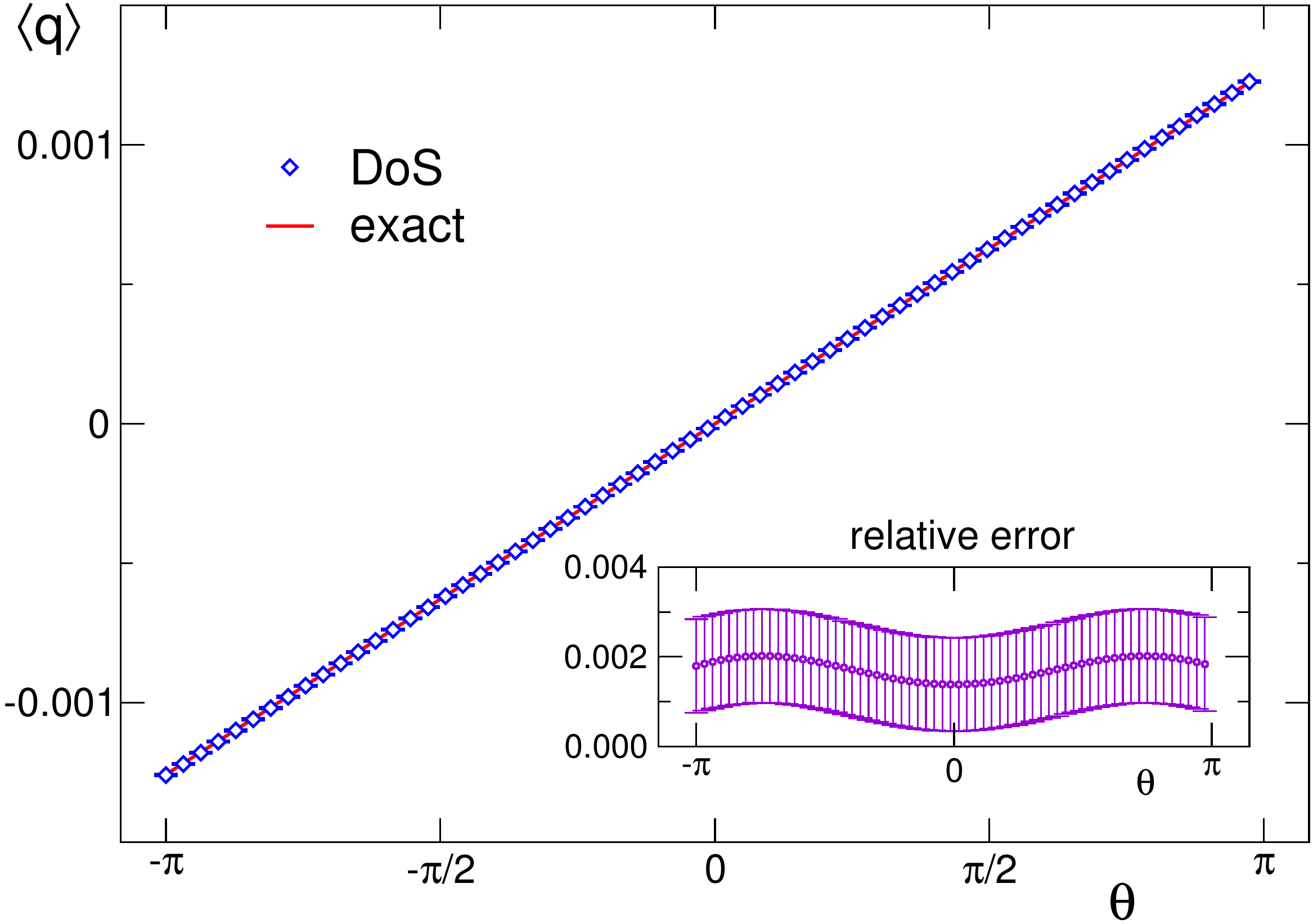}
\caption{Lhs.: Results for the density of states for different volumes. The symbols represent the data from the DoS FFA 
calculation with Wilson gauge action, while the dashed lines are the exact results. For comparison we also display the 
result for the Villain action. Rhs.: Dos FFA results (symbols) for the topological charge density compared to the corresponding 
exact result (full curve). Figures taken from \cite{Gattringer:2020mbf}.}
\label{figure_2}
\end{figure}

Beyond the more fundamental question of  the correct continuum limit, the exact reference results allow one to also address the 
question whether the DoS FFA techniques we use give rise to sufficient accuracy, such that the oscillating integrals (\ref{integrals}) 
can be reliably evaluated. We begin our assessment of the accuracy with the results for the density $\rho(x)$ shown in the 
lhs.\ plot of Fig.~\ref{figure_2}. The data are from a simulation of 2-d U(1) lattice gauge theory with Wilson action and a simple field 
theoretical discretization of the topological charge (see \cite{Gattringer:2020mbf} for the details of the simulation). We show results
for $L \times L$ lattices with different sizes $L$. The inverse gauge coupling $\beta$ was chosen such that the ratio 
$R =  L^2/\beta$ remains constant which means that we approach the continuum limit for a fixed physical volume. In the continuum the 
ratio is given by $R = Ve^2$ where $V$ is the physical volume and $e$ the electric charge. In
Fig.~\ref{figure_2} this ratio is chosen as $R = 10$ and the symbols represent the results from the DoS FFA calculation. The dashed 
lines are the corresponding analytical results and for all volumes we find excellent agreement with the numerical data. For 
comparison we also show the analytical result for the Villain action in the continuum limit at $R = 10$ and the sequence of volumes 
and couplings we consider for the Wilson action at $R = 10$ approaches the corresponding Villain continuum result. It is important to 
note that the densities show a Gaussian-type of shape, i.e., they decay quickly for increasing $x$ such that the truncation 
of $x$ to a finite interval is justified. 

In the rhs.\ plot of Fig.~\ref{figure_2} we show the result for the topological charge density $q = Q/V$
as a function of the topological angle $\theta$. To obtain 
this result one only needs the density $\rho(x)$ of Eq.~(\ref{densities}) for $J \equiv \mathds{1}$, which then is integrated over with 
$e^{-i \theta x}$ to obtain the partition function $Z_\theta$ and with $x \, e^{-i \theta x}$ to obtain the numerator of the vacuum expectation
value $\langle q \rangle$. The corresponding results in the rhs.\ plot of Fig.~\ref{figure_2} are for lattice size $L = 24$, with again $R = 10$. 
We find that the DoS FFA results (symbols) and the corresponding exact results (red full line) match very well for the
full range of $\theta$-values. The insert shows the relative error and we find that it remains below 0.3 \%. 

The tests presented in \cite{Gattringer:2020mbf} show that, at least for the case of 2-d U(1) lattice gauge 
theory with a topological term, the approach of using DoS FFA with open boundary conditions correctly reproduces
the continuum physics at finite $\theta$. Based on analytic results it was shown that the open boundary conditions give
rise to the correct continuum limit, and the numerical tests established that the parameterized density can be obtained with 
sufficient accuracy, such that the necessary integration with the oscillating factors reliably provides vacuum expectation values
at non-zero $\theta$. 

\section{First tests in SU(2) lattice gauge theory}

\noindent
Having built up experience with the application of the DoS FFA to a simple gauge theory with a topological term, and having 
established that in this case the necessary accuracy could be achieved, we now come to the presentation of first results in 
4-d SU(2) lattice gauge theory with a $\theta$-term. Clearly the case of 4-d SU(2) lattice gauge theory brings in new challenges
such as the non-abelian nature of the problem and its higher dimensionality. 

Our calculation is based on the Wilson action and we use the field theoretical definition of the topological charge $Q$ 
\cite{q_field_th_def} ("clover discretization"). We work on lattices of size $16^3 \times 4$, i.e., 
a finite temperature setting and we use mixed boundary 
conditions where one of the spatial directions remains open, while the other space directions and the time direction have 
periodic boundary conditions. Thus the topological charge does not become quantized to integers in the continuum limit and the
density of states remains regular such that we can access it with the DoS FFA. The finite temperature setting, i.e., a short temporal 
direction, has been chosen such that changing the inverse gauge coupling $\beta$ drives the system through the deconfinement 
transition, which allows us to test the DoS FFA approach to in the two different phases of the theory. Besides this technical aspects, 
from a physical point of view it is a first attempt at studying the change of the $\theta$ dependence across the phase transition.

\begin{figure}[t] 
\begin{center}
\includegraphics[scale=0.6,clip]{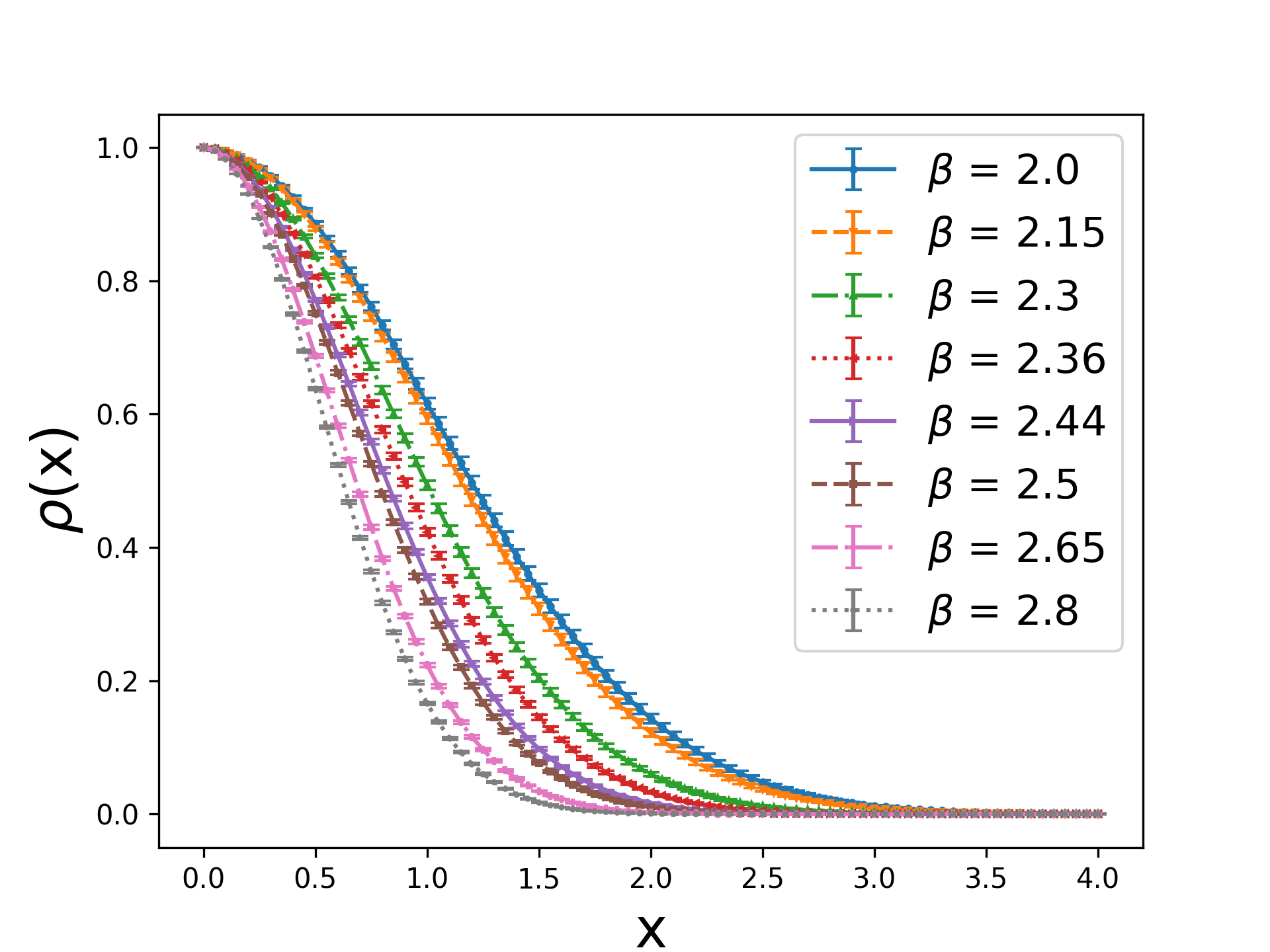} 
\end{center}
\caption{The density of states for 4-d SU(2) lattice gauge theory. The results are for $16^3 \times 4$ lattices and we compare
the results for different values of the inverse gauge coupling $\beta$.}
\label{figure_3}
\end{figure}

The details of the numerical simulation used for the results shown here will be presented in detail in an 
upcoming publication \cite{upcoming}, and we here only address a new technical step that leads to a considerable 
improvement of the results. In Eq.~(\ref{paramdens}) we have written the density in terms of the slopes $k_n$ describing the 
density in the corresponding discretization intervals $I_n$. Using the DoS FFA these $k_n$ are computed for each interval
independently. Clearly the $k_n$ will have statistical errors that will lead to fluctuations around their true values. These fluctuations
will introduce some roughness into the numerical results for the densities, which in turn will spoil the accuracy of the oscillating
integrals (\ref{integrals}) and thus the results for physical observables. However, this naive approach ignores the fact that (at least for
finite lattices) $\rho(x) = e^{-L(x)}$ and thus also the function $L(x)$ in the exponent of the density must be smooth functions. In order 
to take this into account we may make an ansatz for $L(x)$ in the form of an even (or odd) finite polynomial, such as (note 
that we may normalize to $L(0) = 0$, such that no constant term appears in $L(x)$) 
\begin{equation}
L(x) \; \sim \; a_2 \, x^2 + a_4 \, x^4 + \, ... \, a_{2k} \, x^{2k} \; \sim \; \sum_{j = 0}^{n-1} [k_j - k_n] \, \Delta_j \, + \, k_n \, x \quad 
\mbox{for} \quad x \, \in \, I_n  \; ,
\end{equation}
where in the first step we have represented $L(x)$ by the finite polynomial and in the second step exploited (\ref{paramdens})
to write $L(x)$ using our representation in terms of the slopes $k_n$. We may now use the numerical data for all slopes
$k_n$ as input and then determine the values for the coefficients $a_{2j}$ from a global fit to all numerical data. Subsequently
the new parameterized form $\rho(x) = \exp( - a_2 \, x^2 \, ... \, - a_{2k} \, x^{2k})$ is used for evaluating the integrals 
(\ref{integrals}) and thus the observables. Obviously this step, which corresponds to a reparameterization of the density 
now takes into account the smoothness discussed above.

In Fig.~\ref{figure_3} we show the results for the density of states at different values of the inverse gauge coupling $\beta$. 
We remark that for a temporal extent of 4 the critical gauge coupling is $\beta_c \sim 2.35$ \cite{Engels:1988ph}. 
We observe that for $\beta$ above $\beta_c$, i.e., in the deconfined phase, the density is considerably more narrow than 
in the confined phase ($\beta < \beta_c$), which reflects the fact that in the deconfined phase the moments of $Q$ become
suppressed quickly. For the consistency of our approach, in particular the truncation of the $x$-range, it is important to note
that as in the 2-d U(1) case we find that for all values of $\beta$ the densities have a Gaussian-like shape, i.e., they decay
fast, such that the truncation in our parameterization of the densities is justified.  

We conclude our first discussion of the 4-d SU(2) results with an assessment of the accuracy we have achieved in our calculation 
of the density. Clearly no analytical reference results are available in this case, but we may use the density also for the evaluation 
of observables at $\theta = 0$, which can then be compared to results of a conventional simulation. More specifically we look at 
the DoS results for $\langle Q^2 \rangle$ and $\langle Q^4 \rangle$ at $\theta = 0$ and compare them with reference data from a 
conventional simulation. This is done for different inverse gauge couplings, i.e., different temperatures such that we can assess
the accuracy for different physical situations. 

\begin{figure}[t] 
\begin{center}
\hspace*{-2mm} \includegraphics[scale=0.49,clip]{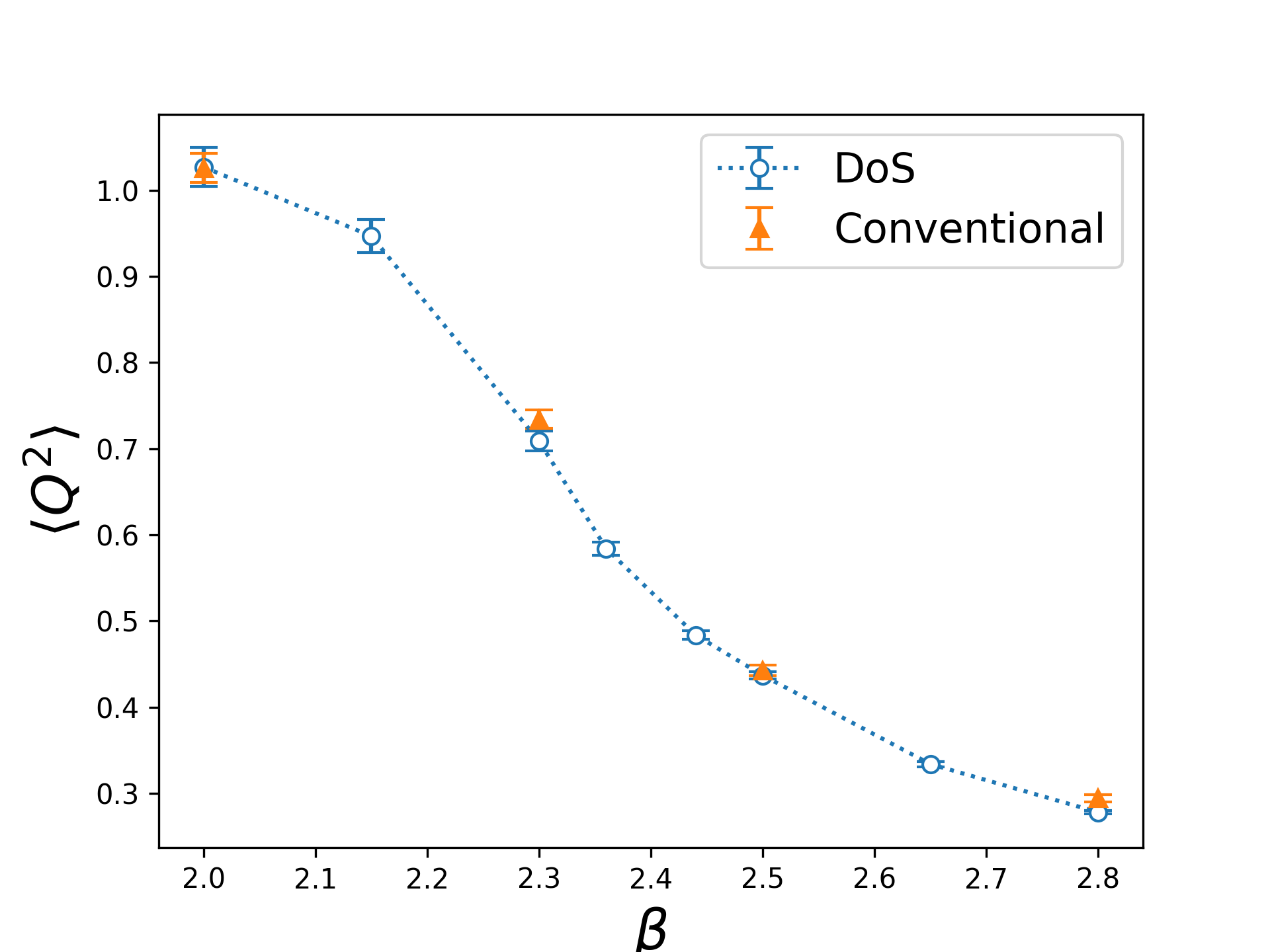} 
\hspace*{-8mm}\includegraphics[scale=0.49,clip]{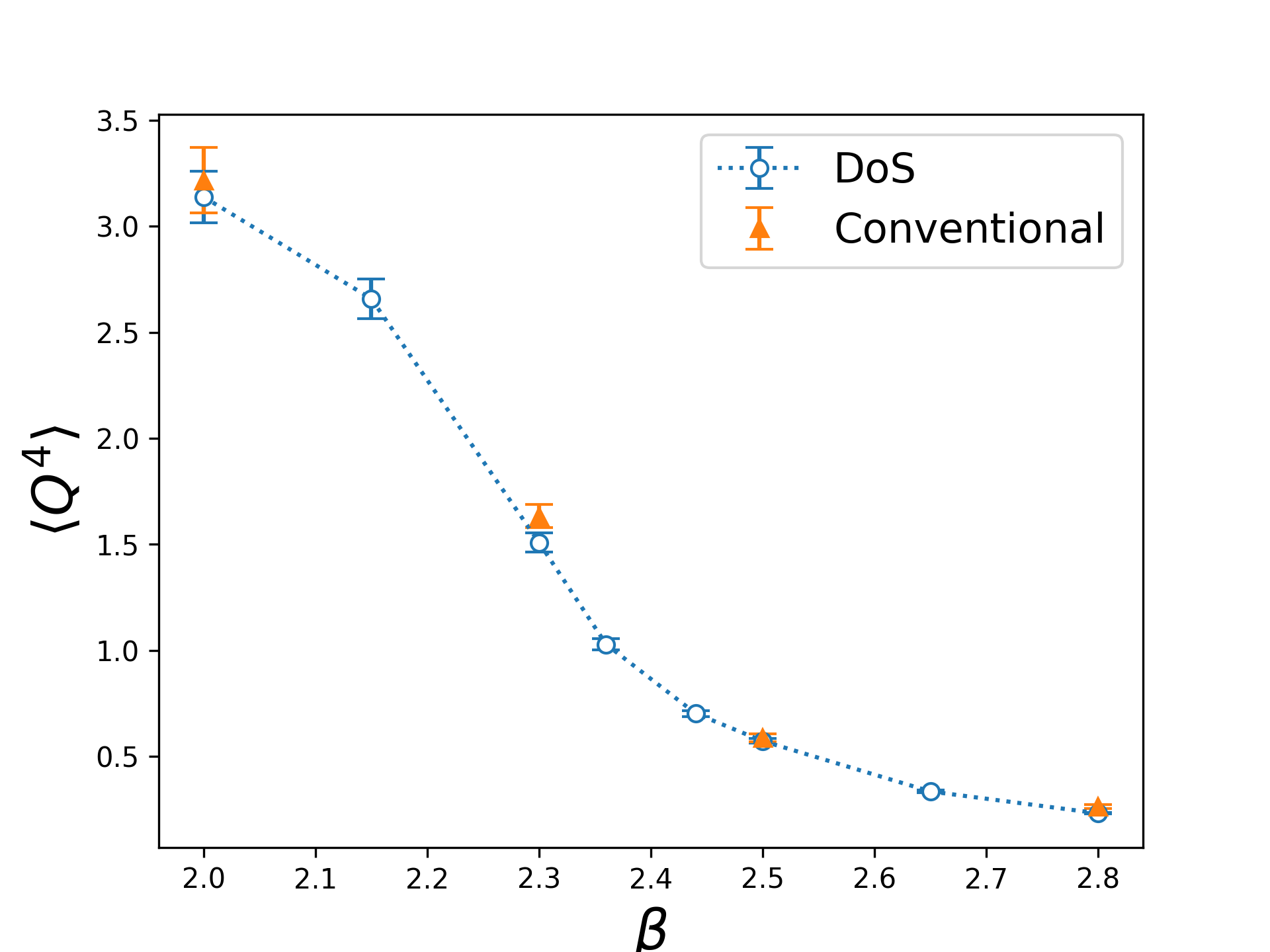}
\end{center}
\caption{Results for $\langle Q^2 \rangle$ (lhs.\ plot) and $\langle Q^4 \rangle$ (rhs.) at $\theta = 0$. 
We show the two moments as a function of the inverse gauge coupling $\beta$ and compare the results 
from the DoS FFA calculation (open circles) to reference data obtained from a conventional calculation at $\theta = 0$ 
(filled triangles).}
\label{figure_4}
\end{figure}

The corresponding results are shown in Fig.~\ref{figure_4} where $\langle Q^2 \rangle$ (lhs.\ plot) and $\langle Q^4 \rangle$ (rhs.)
are shown as a function of $\beta$. The open circles represent the results from DoS FFA, while the filled triangles are the results 
from a conventional simulation. We find very good agreement between the Dos FFA results and the reference data from the 
conventional simulation for all values of $\beta$, which indicates that DoS FFA is ready for the analysis of the $\theta$-dependence
of observables in 4-d SU(2) lattice gauge theory.

\section{Summary and outlook}

\noindent
In this contribution we have reported on our progress with developing density of states techniques for 
lattice field theory with a topological
term. The key ingredient is to use open boundary conditions such that the topological charge is no longer quantized to integers and 
the density has a regular behavior that can be accessed with modern DoS techniques, the DoS FFA in our case. Using 2-dimensional
U(1) lattice 
gauge theory with a $\theta$-term as a simple test case we present results of the method and compare them to analytical results 
available for that model. In particular we also address the question whether our open boundary conditions and the usual periodic choice 
give rise to the same continuum results -- this is confirmed.

In a second part we present our new tests for 4-d SU(2) lattice gauge theory with a $\theta$-term. Here the focus is on
achieving the necessary accuracy for the DoS approach in a theory that is considerably closer to the target theory of QCD. 
We present results for 
the case without topological term and show that the standard simulation results for the second and the fourth moment of the topological 
charge are precisely reproduced by the DoS method in a wide range of temperatures. The method thus is ready for computing 
observables at non-zero topological angle $\theta$ \cite{upcoming}.

\end{document}